\documentclass[preprint,aps,pra,amssymb]{revtex4-2}

\usepackage{graphicx}
\usepackage{dcolumn}
\usepackage{bm}
\usepackage{siunitx}
\usepackage{amssymb}
\usepackage{hyperref}

\newcommand{\Be}{\ensuremath{^9\mathrm{Be}^+\;}}
\newcommand{\BeNoSpace}{\ensuremath{^9\mathrm{Be}^+}}
\newcommand{\Bem}{\ensuremath{^9\mathrm{Be}^-\;}}
\newcommand{\BeBe}{\ensuremath{^9\mathrm{Be}^+ - {^9\mathrm{Be}}^+\;}}

\newcommand{\BemBep}{\ensuremath{^9\mathrm{Be}^- - {^9\mathrm{Be}}^+\;}}

\newcommand{\aHtwomBe}{\ensuremath{\mathrm{\bar{H}}_2^- - {\mathrm{Be}}^+\;}}
\newcommand{\aHtwomBeNoSpace}{\ensuremath{\mathrm{\bar{H}}_2^- - {\mathrm{Be}}^+}}
\newcommand{\HtwopBe}{\ensuremath{\mathrm{H}_2^+ - {\mathrm{Be}}^+\;}}

\newcommand{\Dm}{\ensuremath{\mathrm{D}^-\;}}

\newcommand{\aHtwom}{\ensuremath{\mathrm{\bar{H}}_2^-\;}}
\newcommand{\aHtwomNoSpace}{\ensuremath{\mathrm{\bar{H}}_2^-}}
\newcommand{\Htwop}{\ensuremath{\mathrm{H}_2^+\;}}
\newcommand{\HtwopNoSpace}{\ensuremath{\mathrm{H}_2^+}}
\newcommand{\DmBe}{\ensuremath{\mathrm{D}^- - {\mathrm{Be}}^+\;}}

\newcommand{\kHz}[1]{\qty{#1}{\kilo\hertz}}
\newcommand{\MHz}[1]{\qty{#1}{\mega\hertz}}

\newcommand{\um}[1]{\qty{#1}{\micro\meter}}

\newcommand{\etal}{\textit{et al.\ }}

\begin{document}

\title{Coupling of negative-positive trapped-ion pairs}

\author{Daniel Kienzler}
\email{daniel.kienzler@phys.ethz.ch}
\affiliation{Department of Physics, ETH Zürich, Zürich, Switzerland}
\affiliation{Quantum Center, ETH Zürich, Zürich, Switzerland}

\begin{abstract}
Direct motional coupling of opposite-charge trapped-ion pairs could provide a pathway to extend ultra-low temperatures and quantum control to negative ions that lack the suitable electronic energy structures required for direct laser cooling. Because positive and negative ions cannot be confined within a single electrostatic potential well, I investigate a configuration where single ions are trapped in close proximity within separate potential wells to couple their motion. I analytically and numerically evaluate the electrostatic trapping requirements. As a concrete implementation, I present an optimized segmented surface Paul trap design to couple an antimatter hydrogen molecular ion (\aHtwomNoSpace) and a beryllium ion (\BeNoSpace). A motional coupling frequency of \kHz{5} can be achieved at an ion-ion separation of \um{35}, with an ion height of \um{50}, axial trap frequencies of \MHz{4}, and static trap voltages with a magnitude of $\approx\qty{20}{\volt}$.
Finally, I outline three applications for this technique: quantum logic spectroscopy of \aHtwom for matter-antimatter comparisons, the preparation of cold neutral deuterium atoms via near-threshold photo-detachment of \Dm for optical trapping, and quantum information processing using equal-mass opposite-charge ion pairs.
\end{abstract}

\maketitle

\section{Introduction}
    \noindent Progress in atomic and molecular physics has been strongly driven by laser cooling and trapping. Laser cooling however, requires a suitable electronic energy structure. To extend ultra-low temperatures to species not suitable for laser cooling, sympathetic cooling has been utilized \cite{86Larson,1997Myatt,1999DeMarco,2001Bloch,2001Modugno,2001Schreck,03Barrett}. For ions, this is performed by co-trapping a coolable species with the species of interest, relying on their mutual Coulomb interaction to extract energy \cite{86Larson,03Barrett}. This principle has been extended in quantum logic spectroscopy to additionally enable quantum control of an ion of interest \cite{05Schmidt}, and is today applied to a wide variety species: single-ionized atoms, highly-charged ions, and molecular ions \cite{2025Marshall,2020Micke,2022King,16Wolf,17Chou,2020Sinhal,2025Holzapfel}. 
    
    So far laser cooling has only been demonstrated with positively charged ions. Most negatively charged ions do not possess a bound excited state required for laser cooling, and only few species with the potential for laser cooling have been identified thus far \cite{2009Warring,2014Walter,2015Yzombard,2018Cerchiari,2022Notzold,2025Zhang}. This might suggest sympathetic cooling of the negative ion species with laser-cooled positive ions, but this approach is hampered by the fact that positive and negative ions cannot be confined in the same electrostatic potential well. 
    A charge-sign agnostic sympathetic cooling method has been demonstrated with protons, implemented by coupling the motion of laser-cooled \Be ions through image currents \cite{2021Bohman,2024Will}. 
    
    A different approach is to trap positive and negative ions in individual potential wells, but in close proximity to couple their motion directly. 
    This has been proposed for cooling and implementing quantum logic readout of the (anti-)proton \cite{98Wineland1}. Recently, numerical studies for cooling and coupling antiprotons and anti-hydrogen molecular ions to \Be ions at large distances have been presented ($\mathcal{O}(\um{100})$) \cite{2026Poljakov,2026Schiller}.
    
    Here I follow this approach and study the details of negative-positive ion coupling in the limit of single ions and close distance ($\mathcal{O}(\um{10})$) with the goal of implementing motional ground-state cooling and quantum logic spectroscopy. 
    I discuss three applications for this technique: 1. Implementing quantum logic spectroscopy of negatively charged species such as the antimatter hydrogen molecular ion \aHtwomNoSpace, which would enable comparisons to its matter counterpart \Htwop \cite{18Myers,2023Schiller,2025Alighanbari,2025Holzapfel,2026Schiller}. 2. A source of cold neutral atomic and molecular species which cannot be laser cooled efficiently, by cooling their negative ions and performing photo-detachment close to the threshold. 3. The coupling of same-mass negative-positive ion pairs for quantum information processing.
        
    Section \ref{sec:Trapping potentials and trap geometries} of this letter  discusses the requirements for the trapping potential and details of the negative-positive-ion coupling and presents an optimized segmented ion trap design. Section \ref{sec:applications} discusses potential applications.

\section{Trapping potentials and trap geometries} \label{sec:Trapping potentials and trap geometries}
    \subsection{Theoretical description} \label{sec:theoretical_description}
        \noindent The strength of the Coulomb interaction between two trapped ions close to their equilibrium positions can be expressed as the frequency of energy exchange given by \cite{11Brown1}
        \begin{eqnarray}
            \Omega_{\mathrm{ex}} = \frac{q_a q_b}{4 \pi \epsilon_0 d_{\mathrm{equi}}^3 \sqrt{m_a m_b} \sqrt{\omega_a \omega_b}},
        \end{eqnarray}
        with the ions labeled $a$ and $b$ and their masses $m_{a/b}$, charges $q_{a/b}$, trap frequencies $\omega_{a/b}$, ion-ion equilibrium distance $d_{\mathrm{equi}}$, and the vacuum permittivity $\epsilon_0$.
        Thus, to achieve a high coupling strength, the ions need to be brought into proximity. 
        For an ion pair of the same charge, this can be easily achieved by confining them in a common, harmonic potential well, or by confining them in symmetric, separate potential wells \cite{11Brown1, 14Wilson}.
        For a negative-positive ion pair it is not possible to confine the particles in a common electro-static potential well: a confining well for a positively charged particle creates anti-confinement for a negative particle and vice versa. Additionally, confining negative and positive ions in a common charge-agnostic potential as a point-Paul trap provides, would lead to charge exchange reactions, limiting the ions' trapping lifetime. 
        To confine a positive-negative ion pair in separate potential wells, an anti-symmetric trapping potential is required to create confinement for the different charges. Additionally, a sufficiently strong potential barrier between them is required to overcome their Coulomb attraction, preventing the ion pair from colliding. 
        
        The problem is analyzed in one spatial dimension ($z$), relying on a static electric potential for confinement. For this initial analysis the ion-ion Coulomb interaction is neglected. It is then added in sections \ref{sec:Equal masses} and \ref{sec:Unequal masses}.
        We aim to form an electric potential $\phi$ with turning points at $z = \pm \delta/2$ and curvatures $\partial^2\phi/\partial z^2|_{\pm \delta/2} = \omega_{\pm}^2 m_{\pm} / q_{\pm}$, where $\omega_{\pm}$ are the trap frequencies (without ion-ion Coulomb interaction) for particles with the masses $m_{\pm}$ and charges $q_{\pm}$ at the locations $\pm \delta/2$. The imposed symmetry around the origin does not limit the solution but simplifies the mathematical expressions.
        The lowest-order solution is given by \begin{eqnarray}\label{eq:general_potential}
            \phi \left( z \right) =  \alpha_1 z + \alpha_2 z^2 + \alpha_3 z^3 + \alpha_4 z^4
        \end{eqnarray}
        with coefficients
        \begin{eqnarray}
        \alpha_1 &=& \frac{\delta}{8} \left(\frac{m_- 
        \omega_-^2}{q_-} - \frac{m_+ \omega_+^2}{q_+} \right),\\
        \alpha_2 &=& -\frac{1}{8} \left(\frac{m_- 
        \omega_-^2}{q_-} + \frac{m_+ \omega_+^2}{q_+} \right),\\
        \alpha_3 &=& - \frac{1}{6 \delta} \left(\frac{m_- 
        \omega_-^2}{q_-} - \frac{m_+ \omega_+^2}{q_+} \right),\\
        \alpha_4 &=& \frac{1}{4 \delta^2} \left(\frac{m_- 
        \omega_-^2}{q_-} + \frac{m_+ \omega_+^2}{q_+} \right).
        \end{eqnarray}
        
\begin{figure}
            \centering\resizebox{0.48 \textwidth}{!}{
            \includegraphics{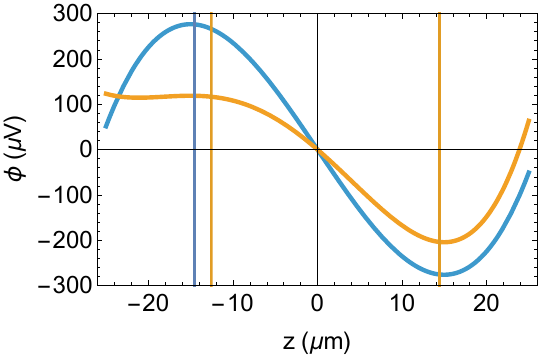}}
            \caption{Thick curves are electric potentials for trapping positive and negative ions along the $z$ direction. For the blue potential curve the parameters are $\delta=\qty{30}{\micro\meter}$, $\omega_{\pm}= 2\pi \times \qty{1}{\mega\hertz}$, $m_{\pm}=\qty{9}{\dalton}$, for the orange potential curve $m_{+}=\qty{9}{\dalton}$, $m_{-}=\qty{2}{\dalton}$ and all other parameters as for the blue curve. The vertical thin lines mark the equilibrium positions of the coupled ions (left negatively, right positively charged) for the potentials plotted in the same color.}
            \label{fig:potentials}
        \end{figure}

        It is notable that for an anti-symmetric potential ($m_+ \omega_+^2/q_+ = -m_- \omega_-^2/q_-$) it follows that $\alpha_2, \alpha_4 = 0$. This is fulfilled for a negative-positive ion pair of equal masses and charges at the same trap frequencies. In contrast, for a symmetric potential ($m_+ \omega_+^2/q_+ = m_- \omega_-^2/q_-$) which might be used for two positive ions of the same mass, $\alpha_2, \alpha_4 \ne 0$ but $\alpha_1, \alpha_3 = 0$.
        However, resonant coupling ($\omega_+\approx\omega_-$) of an ion pair with different masses will require $\alpha_1, \alpha_3, \alpha_2, \alpha_4 \ne 0$. 
        Example potential shapes are shown in figure \ref{fig:potentials}. 
        
        \subsection{Equal masses: \Be and \Bem} \label{sec:Equal masses}
        To describe the basic properties of the positive-negative ion pair coupling I first discuss the case of identical masses ($m = m_{\pm}$) and trap frequencies ($\omega = \omega_\pm$) and opposite charges ($q = q_+ = -q_-$), resulting in an anti-symmetric potential ($\alpha_2 = \alpha_4 = 0$). For this configuration analytic expressions can be found easily. Numerical results are given for a fictitious \BemBep pair (as the \Bem anion is not stable).
        We define $\xi \equiv m \omega^2/q$, resulting in $\alpha_1 = - \xi \delta/4$ and $\alpha_3 = \xi/(3 \delta)$.
        The potential energy for the ions at symmetric distance $d/2$ from the origin ($z_+ = d/2$, $z_- = -d/2$), including their Coulomb interaction, is then given by 
        \begin{eqnarray}
            U = 2 q \phi(d/2) - \frac{q^2}{4 \pi \epsilon_0 |d|}.
        \end{eqnarray}
        The equilibrium distance $d_\mathrm{equi}$ of the coupled ions can be found by minimizing $U$ with respect to $d$, resulting in 
        \begin{eqnarray}
            d_\mathrm{equi} = \sqrt{\frac{\delta^2 + \sqrt{\frac{\delta}{\pi \epsilon_0 \xi}(\delta^3 \pi \epsilon_0 \xi - 4 q)}}{2}}.
        \end{eqnarray}
        
        If $\delta$ is chosen too small, the confining potential is unable to hold the ions in their wells and they collide. The critical equilibrium distance $d_{\mathrm{equi, c}}$ and the corresponding $\delta_\mathrm{c}$ are found by setting the inner radicand to zero ($\delta^3 \pi \epsilon_0 \xi - 4 q = 0$), resulting in
        \begin{eqnarray}
           d_{\mathrm{equi, c}} =  \left( \frac{\sqrt{2} q}{\pi \epsilon_0 \xi}\right)^{1/3} \quad \mathrm{and} \quad \delta_\mathrm{c} = \sqrt{2} d_{\mathrm{equi, c}}.
        \end{eqnarray}

\begin{figure*}
            \centering
            \includegraphics[width=\textwidth]{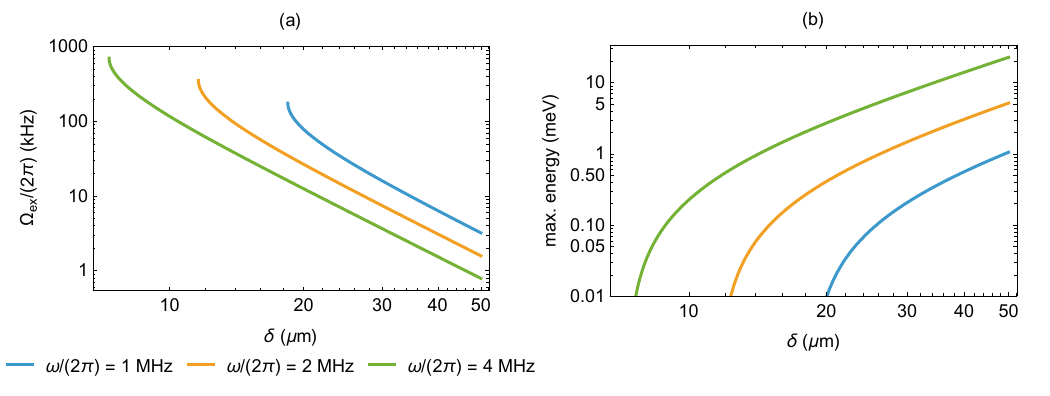}
            \caption{(a) Coupling strength $\Omega_{\mathrm{ex}}$ and (b) maximum stable energy as a function of the turning point distance $\delta$ for the \BemBep ion pair ($m_{\pm} = \qty{9}{\dalton}$). The blue, orange, green curves are for $\omega_{\pm}/(2 \pi) = 1,2,\qty{4}{\mega\hertz}$, respectively.}
            \label{fig:bebe}
        \end{figure*}

        In an experimental setting, $\delta$ has to be chosen sufficiently large to achieve stable trapping of ions with finite energy but small enough to achieve the desired coupling. For experimentally convenient operation we are interested in achieving a coupling rate of $\Omega_{\mathrm{ex}}/(2\pi) \gtrsim \qty{5}{\kilo\hertz}$, similar to Wilson \etal \cite{14Wilson}.
        An estimate for the maximum stable energy of the ions for a given potential can be found from the potential energy difference at the turning points ($\partial U/ \partial d = 0$). This results in long expressions and is thus given as a numerical result along with the coupling strength $\Omega_{\mathrm{ex}}$ in figure \ref{fig:bebe}.
        E.g.\ a maximum stable energy of \qty{6}{\milli\electronvolt} and a coupling strength of $\Omega_{\mathrm{ex}}/(2\pi) = \qty{5}{\kilo\hertz}$ are achieved for $\omega_{\pm} / (2\pi) = \qty{4}{\mega\hertz}$, $\delta = \qty{27}{\micro\meter}$.

        The normal modes of motion for the coupled system can be found by numerically calculating the Hessian and its eigensolutions at $\bar{z}_{\pm}$ \cite{13Home}. Results for the axial modes of the \BemBep pair are given in table \ref{tab:normal_modes}. The ions oscillate in and out of phase with two normal mode frequencies. Some differences to the usually studied two-ion same-charge-sign coupling can be observed: The normal modes have lower frequencies than the uncoupled motion as they come closer together and thus experience a lower potential curvature; the in-phase mode is the higher frequency and the out-of-phase mode is the lower frequency mode; the in-phase mode can only be excited by an electric field gradient (oscillating with the in-phase mode frequency), the out-of-phase mode by a homogeneous field (oscillating with the out-of-phase mode frequency). All these facts are a result of the attraction of the two ions (instead of the usually encountered repulsion).

        An important experimental limitation in coupling the motion of ion pairs is the impact of stray electric fields. For a same-charge-sign ion pair a homogeneous stray field pushes both ions into the same direction. This leads to a lower trap frequency for the ion pushed towards the center ($z=0$) as it experiences weaker confinement, and a higher trap frequency for the ion pushed away from the center  as it experiences stronger confinement. This shifts the ions' motional frequencies out of resonance, compromising the desired coupling.
        For positive-negative ion pairs a homogeneous stray field pushes the ions into opposite directions, leading to a symmetric change in the trap frequency and thus preserving the resonance condition while still changing the normal mode frequencies.
        To characterize the sensitivity to stray fields it is instructive to evaluate the change in normal mode frequency and mode participation due to an applied electric field. The results are presented in table \ref{tab:normal_modes} and confirm the described behavior: While the trap frequencies shift for the \BemBep pair, the ions' mode participation stays equal for both ions in both axial modes. This makes the negative-positive ion pair with equal masses a more robust system compared to same-charge-sign ion pairs.

        \begin{table}
        \caption{\label{tab:normal_modes}
        Normal modes of example ion pairs. The in- and out-of-phase mode frequencies $\omega_{\mathrm{in/out}}$ and the normal mode amplitude vectors for the individual ions ($+,-$) $\bm{e}_{\mathrm{in/out}}\{+,-\}$ are determined from numerically calculating the Hessian and its eigensolutions. For the \aHtwomBe (\HtwopBe) ion pair $\omega_- = 2\pi \times \qty{3.978}{\mega\hertz}$ ($\omega_- = 2\pi \times \qty{3.930}{\mega\hertz}$) is chosen to achieve resonant coupling. An electric field $E$ in the axial direction is added to the ideal potentials to elucidate the sensitivity of the ion pairs to stray fields. For comparison, results for the \BeBe and \HtwopBe ion pairs are included. For \HtwopBe $\omega_-$ refers to the frequency of \HtwopNoSpace. The negative-positive ion pairs show a lower sensitivity to stray electric fields in both the normal mode frequency shift and the mode participation compared to their positive-positive counterparts.}
        
        \centering
        \begin{tabular}{lllllllllll}
        \hline
        \textrm{Ion pair} &
        $\delta$ &
        $\omega_{\pm}/(2\pi)$&
        $E$ &
        $\omega_{\mathrm{in/out}}/(2\pi)$&
        $\bm{e}_{\mathrm{in}}\{+,-\}$&
        $\bm{e}_{\mathrm{out}}\{+,-\}$\\
         &
        (\unit{\micro\meter})&
        (MHz)&
        (V/m) &
        (MHz)&
        &
        \\
        \hline
        \BeNoSpace,\Bem & 25 & 4.000, 4.000 & 0 & 3.995, 3.985 & $\{0.71,0.71\}$ & $\{0.71,-0.71\}$\\
        \BeNoSpace,\Bem & 25 & 4.000, 4.000 & 1 & 3.992, 3.982 & $\{0.71,0.71\}$ & $\{0.71,-0.71\}$\\
        \BeNoSpace,\Bem & 25 & 4.000, 4.000 & -1 & 3.998, 3.988 & $\{0.71,0.71\}$ & $\{0.71,-0.71\}$\\
        \hline
        \BeNoSpace,\Be & 25 & 4.000, 4.000 & 0 & 4.015, 4.025 & $\{0.71,0.71\}$ & $\{0.71,-0.71\}$\\
        \BeNoSpace,\Be & 25 & 4.000, 4.000 & 1 & 4.011, 4.029 & $\{0.29,0.96\}$ & $\{0.96,-0.29\}$\\
        \BeNoSpace,\Be & 25 & 4.000, 4.000 & -1 & 4.011, 4.029 & $\{0.96,0.29\}$ & $\{0.29,-0.96\}$\\
        \hline
        \BeNoSpace,\aHtwom & 35 & 4.000, 3.978 & 0 & 3.999, 3.989 & $\{0.71,0.71\}$ & $\{0.71,-0.71\}$\\
        \BeNoSpace,\aHtwom & 35 & 4.000, 3.978 & 1 & 4.016, 3.989 & $\{0.98,0.19\}$ & $\{0.19,-0.98\}$\\
        \BeNoSpace,\aHtwom & 35 & 4.000, 3.978 & -1 & 3.998, 3.971 & $\{0.18,0.98\}$ & $\{0.98,-0.18\}$\\
        \hline
        \BeNoSpace,\Htwop & 35 & 4.000, 3.930 & 0 & 4.002, 4.012 & $\{0.71,0.71\}$ & $\{0.71,-0.71\}$\\
        \BeNoSpace,\Htwop & 35 & 4.000, 3.930 & 1 & 4.003, 4.063 & $\{1.00,0.08\}$ & $\{0.08,-1.00\}$\\
        \BeNoSpace,\Htwop & 35 & 4.000, 3.930 & -1 & 3.950, 4.012 & $\{0.08,1.00\}$ & $\{1.00,-0.08\}$\\
        \hline
        \end{tabular}
        \end{table}

        \subsection{Unequal masses: \Be and \aHtwom} \label{sec:Unequal masses}
\begin{figure*}
            \centering
            \includegraphics[width=\textwidth]{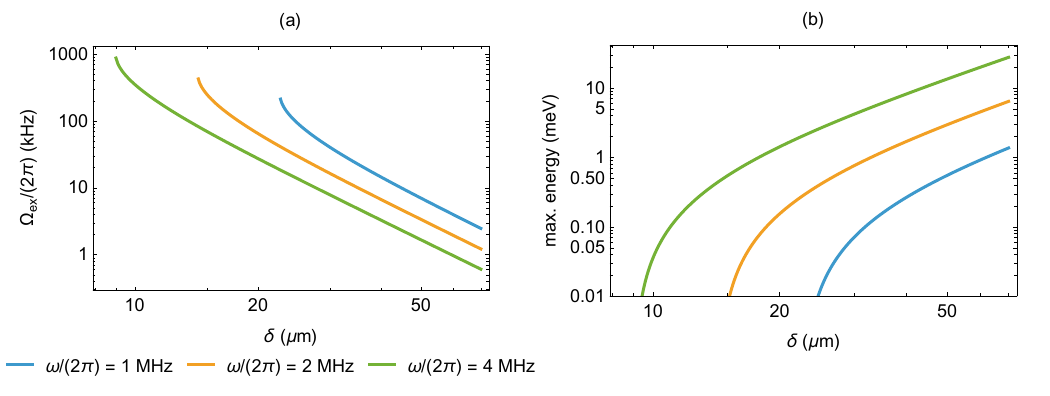}
            \caption{(a) Coupling strength $\Omega_{\mathrm{ex}}$ and (b) maximum stable energy as a function of the turning point distance $\delta$ for $m_{+} = \qty{9}{\dalton}$, $m_{-} = \qty{2}{\dalton}$. The blue, orange, green curves are for $\omega_{\pm}/(2 \pi) = 1,2,\qty{4}{\mega\hertz}$, respectively.}
            \label{fig:beh2}
        \end{figure*}

        For unequal masses of the ion pair we study the example of a negative ion with a mass of \qty{2}{\dalton} like the deuterium anion \Dm or the antimatter hydrogen molecular ion \aHtwomNoSpace combined with a positive \Be ion. An example of the  asymmetric potential required for resonant coupling of their axial motion with $\alpha_{1,2,3,4}\ne0$ is given in figure \ref{fig:potentials}. While the potential is very shallow as the negative ion moves away from the positive ion, the maximum stable energy is still limited by the ion-pair attraction. I evaluate the normal modes (table \ref{tab:normal_modes}), and the maximum stable energy and coupling strength (figure \ref{fig:beh2}) for this ion pair numerically.
        Due to the asymmetry of the potential, choosing identical values for uncoupled trap frequencies, e.g.\ $\omega_{\pm}/(2\pi) = \qty{1}{\mega\hertz}$ does not result in fully resonant coupling and requires fine-tuning of $\omega_{\pm}$. For the normal mode analysis, this was performed. For the evaluation of the coupling strength and the maximum stable energy this was omitted for simplicity, but should alter the results only to a minor degree. As can be seen from the figures, e.g.\ $\omega_{\pm}/(2\pi) = \qty{4}{\mega\hertz}$ and $\delta=\qty{35}{\micro\meter}$ result in a coupling strength of $\Omega_{\mathrm{ex}}/(2\pi) = \qty{5}{\kilo\hertz}$ and a maximum stable energy of \qty{6}{\milli\electronvolt}, providing reasonable experimental parameters.

        The maximum stable energy can be compared to the average thermal energy of a \Be ion cooled to the Doppler limit, which is \qty{26}{\nano\electronvolt}. Reaching the Doppler limit for the negatively charged ion by sympathetic cooling will require pre-cooling at larger ion-ion distances. This regime has recently been numerically studied for combining \Be with the antiproton and also the anti-hydrogen molecular ion \cite{2026Poljakov,2026Schiller}. Since the radial motion of two ions of different mass only couples weakly, parametric coupling of the negative ions' radial motion to the axial motion will likely be required to cool it. Such parametric coupling has been demonstrated for mixed-species ion pairs \cite{2024Hou,2025Holzapfel} and should not pose a fundamental issue.

        The protection from stray electric fields displayed by the equal mass pairs is deteriorated for \aHtwomBeNoSpace, due to the asymmetry of the confining potential. However, it still displays a lower sensitivity when compared to the positive-positive \HtwopBe pair (see table \ref{tab:normal_modes}). The sensitivity of \aHtwomBeNoSpace to stray fields is approximately a factor two higher than for the positive-ion \BeBe pair. Since Wilson \etal \cite{14Wilson} have demonstrated the \BeBe coupling with similar parameters, it should be feasible to operate with \aHtwomBeNoSpace. 
        
        Operating with a mass ratio less extreme than \aHtwomBe partially recovers the protection from stray fields visible for identical masses.
        
        \subsection{Three-dimensional confinement, optimized trap geometry}
        Three-dimensional trapping requires further confinement in the radial $x,y$ directions, while leaving the static electric axial confinement untouched. 
        
        This can be achieved by an rf pseudopotential of a linear Paul trap, which provides confinement independent of the charge sign of a particle. The rf pseudopotential overpowers the radial anti-confinement caused by the static potential. This solution offers a convenient choice, as the pseudopotential can be chosen strong enough for the exact form of the radial anti-confining potential to become irrelevant, requiring only to tailor the axial  ($z$) part of the static potential.

        Alternatively, a Penning trap can be used. For this a homogeneous magnetic field along the $z$ direction combined with an anti-confining electric potential in the $x,y$ direction would provide radial confinement. The required static electric anti-confinement can be created without altering the axial static potential while fulfilling Laplace's law. For a positive ion the Penning trap requires an electric potential with a negative curvature in the radial directions. Due to Laplace's law, the axial positive curvature for confinement causes a negative curvature in at least one radial direction, and allows for a radially symmetric negative curvature, which would fulfill the requirement for the Penning trap. For a negative ion the situation is inverted, but results in the same: a negative curvature axially provides axial confinement and causes a positive curvature radially, providing the required electric anti-confinement radially. For robust trapping the anti-confinement should be radially symmetric, which will require optimization of the electrode geometry and electrode voltages. 
        
\begin{figure*}
            \centering
            \includegraphics[width=.6 \textwidth]{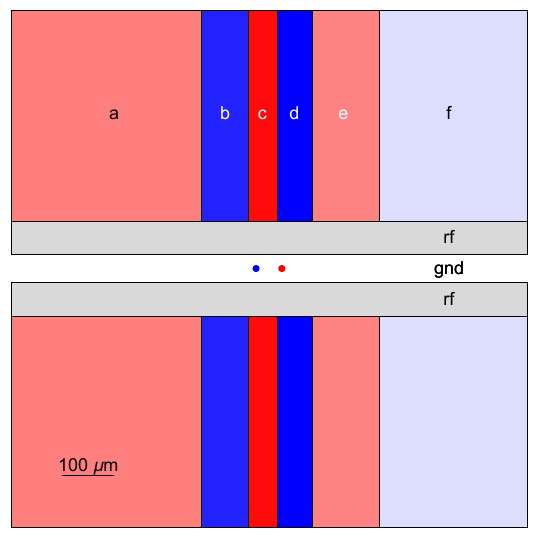}
            \caption{Top view of a surface Paul trap geometry optimized for \aHtwomBeNoSpace. The ion pair is confined in the center (blue dot: \aHtwomNoSpace, red dot \BeNoSpace) \qty{50}{\micro\meter} above the surface with $\omega_{\pm}/(2\pi) = \qty{4}{\mega\hertz}$ and $\delta = \qty{35}{\micro\meter}$. The widths of the DC electrodes are optimized to require the lowest voltage to create the potential. The DC electrodes a-f and their symmetric counterparts below the rf electrodes are biased at a: \qty{11.4}{\volt}, b: \qty{-19.5}{\volt}, c: \qty{21.5}{\volt}, d: \qty{-22.4}{\volt}, e: \qty{11.0}{\volt}, f: \qty{-3.0}{\volt} and have widths a: \qty{369}{\micro\meter}, b: \qty{91}{\micro\meter}, c: \qty{55}{\micro\meter}, d: \qty{68}{\micro\meter}, e: \qty{131}{\micro\meter}, f: \qty{286}{\micro\meter}. The rf electrodes and the center ground electrode have a width of 65 and \qty{54}{\micro\meter}, respectively.}
            \label{fig:trap}
        \end{figure*}

        In either case, due to the strongly anharmonic potential shapes, micro-traps will be required for the implementation. As an example I optimize the design of a linear surface Paul trap geometry for the \aHtwomBe ion pair with $\omega_{\pm}/(2\pi) = \qty{4}{\mega\hertz}$, $\delta = \qty{35}{\micro\meter}$, and an ion height of \qty{50}{\micro\meter}. The required static voltages to create the asymmetric potential are in the experimentally well accessible range of 21.5 to \qty{-22.4}{\volt}. The trap geometry is shown in figure \ref{fig:trap}. A symmetric axial potential with up to fourth order terms requires four electrode segments on either side of the rf electrodes. Here, the potential asymmetry and the requirement to create a field null at both ion positions, require six electrode pairs.

\section{Applications}\label{sec:applications}

    \subsection{Quantum logic spectroscopy of the anti-hydrogen molecular ion}
        The antimatter hydrogen molecular ion has been proposed as a promising candidate for future matter-antimatter comparisons \cite{95Dehmelt,18Myers} and proposals for its synthesis have been made \cite{2025Zammit}. If an anti-hydrogen molecular ion \aHtwom can be co-trapped with a \Be ion as described in section \ref{sec:Unequal masses}, quantum logic spectroscopy, similar to that performed by Holzapfel \etal \cite{2025Holzapfel} could be performed. This would enable high-precision spectroscopy of \aHtwomNoSpace. Spectroscopy of selected rovibrational transitions in \Htwop is projected to achieve fractional frequency uncertainties as low as $\sim10^{-17}$ and \Htwop has been proposed as an optical molecular clock \cite{14Schiller,14Karr}. Using quantum logic spectroscopy, similar precision might be achievable for \aHtwomNoSpace, enabling high-precision matter-antimatter comparisons and the implementation of an optical clock based on antimatter.
        
        Since \aHtwom has not yet been synthesized and will likely only be produced in low quantities, a convenient placeholder for studying the negative-positive ion pair coupling would be the \DmBe ion pair.
        
    \subsection{Preparation of cold deuterium atoms for optical trapping}
        Many atomic species, like hydrogen, cannot be laser cooled efficiently, making it challenging to optically trap them. I here propose instead to sympathetically cool a negative ion by coupling it to a positive ion, followed by an optical electron-detachment of the negative ion close to threshold and capturing of the now neutral atom in an optical dipole trap. 
        This idea is based on schemes originally developed for anti-hydrogen production, namely within the GBAR collaboration for free-fall experiments \cite{2015Perez} and by Crivelli and Kolachevsky for optical trapping \cite{2020Crivelli}.
        
        One application of this scheme would be the preparation of cold deuterium atoms. By coupling a \Dm ion to a \Be ion in the potential discussed in section \ref{sec:Unequal masses}, ground-state cooling of the \Dm ion's motion should be achievable. Subsequent photo-detachment close to threshold would create a cold D atom, which could be optically trapped. This approach could improve high-precision spectroscopy of e.g.\ the deuterium 1S-2S transition by suppressing motional shifts and provide an alternative route towards cold hydrogen isotopes and a calculable clock \cite{2022Vazquez-Carson,2022Jones,2024Amit,2025Amit}.

        The scheme could also be applied to many atomic and molecular species if low temperature e.g.\ for optical trapping is of interest and cannot be achieved by direct laser cooling of the neutral species. The main requirement is that its negative ion can be created.

    \subsection{Quantum information processing}
        As discussed in section \ref{sec:theoretical_description}, the coupling of the motion of a negative-positive ion pair of equal mass can be performed with a potential up to third order, where a fourth-order potential is required for same-charge ion pairs. Thus for achieving the same coupling frequency $\Omega_{\mathrm{ex}}$, this would enable to operate with larger distances from trap electrodes for negative-positive ion pairs compared to same-charge ion pairs. Since the motional heating rate and trap frequency drifts typically scale with electrode-ion distance, this could provide a benefit in performing high-fidelity entangling operations, such as demonstrated by Wilson \etal \cite{14Wilson}. An additional benefit is the protection from stray electric fields that positive-negative ion pairs exhibit (see section \ref{sec:Equal masses}).  
        A challenge for this approach is the lack of obviously suitable negative ion candidates. A starting point for experimental investigation might be the similar-mass  $^{40}\mathrm{Ca}^+ - {^{37}\mathrm{Cl}}^-$ ion pair.

\section{Conclusion}
    I have presented the fundamentals of coupling negative-positive ion pairs. Key sensitivities and implementation parameters are more advantageous than the demonstrated coupling of positive ion pairs \cite{11Brown1,14Wilson}, underscoring the feasibility of coupling positive-negative ion pairs. The approach introduces a novel method for controlling negative ions, offering the potential to reach ultra-cold temperatures and facilitate quantum logic between positive and negative ions. Such a toolbox would provide a pathway to implement quantum logic spectroscopy for the anti-hydrogen molecular ion, enable the preparation of ultra-cold atoms and molecules that are difficult to laser cool, and serve as a new avenue in quantum information processing with trapped ions.
    
\begin{acknowledgments}
The author thanks P. Crivelli, J. P. Home, and F. Schmid for stimulating discussions and careful reading of the manuscript, and the Negative Ion project students (S. Koch, S. Bibawi, S. Agrawal, N. Schwegler, A. Fyrillas, G. Tomassi, M. Simoni, R. Nejad, G. Engin-Deniz) for early contributions on the topic.

This study was supported by ETH Z\"urich and the Swiss National Science Foundation (grant number 190630).
\end{acknowledgments}

\bibliographystyle{iopart-num}
\bibliography{refs}

\end{document}